\documentclass[prb,twocolumn,showpacs,preprintnumbers,amsmath,amssymb]{revtex4}

\usepackage{graphicx}
\usepackage{dcolumn}
 
\newcommand{\beq}{\begin{equation}} 
\newcommand{\eeq}{\end{equation}} 
\newcommand{\beqa}{\begin{eqnarray}} 
\newcommand{\eeqa}{\end{eqnarray}}

\newcommand{\e}{\epsilon}

\begin{document} 
\title{Temperature dependence of the optical spectral weight in the cuprates: 
Role of electron correlations}
\author{A. Toschi$^{1,2}$,  M. Capone$^{1,2,3}$, M. Ortolani$^{1,4}$, P. Calvani$^{1,4}$, S. Lupi$^{1,4}$ and C. Castellani$^{1,2}$}  
\affiliation{$^1$ 
Dipartimento di Fisica, Universit\`a di Roma ``La Sapienza'',
Piazzale Aldo Moro 2, I-00185 Roma, Italy}
\affiliation{$^2$ SMC-INFM}
\affiliation{$^3$ Istituto dei Sistemi Complessi (ISC) del CNR, Via dei Taurini 19, 00185, Roma, Italy} 
\affiliation{$^4$ "Coherentia"- INFM}

\begin{abstract}
We compare calculations based on the Dynamical Mean-Field Theory of the Hubbard model with the infrared spectral weight $W(\Omega,T)$ of La$_{2-x}$Sr$_x$CuO$_4$ and other cuprates. Without using fitting parameters we show that most of the anomalies found in $W(\Omega,T)$ with respect to normal metals, including the existence of two different energy scales for the doping- and the $T$-dependence of $W(\Omega,T)$, can be ascribed to strong correlation effects. 
\end{abstract}

\date{\today} 
\maketitle 

The rich and complex optical response of high-temperature superconducting cuprates (HTS) is still attracting broad interest in the scientific community. In particular, much recent work, both experimental\cite{vdm,bont,homes,boris,michele} and theoretical\cite{hirsh,norman,millis1,carbotte,lara,millis2}  was focused on the temperature dependence of the Cu-O plane infrared spectral weight 

\begin{equation}
W(\Omega,T)=\int_0^{\Omega} d\omega \sigma_1(\omega).
\label{weight}
\end{equation}

\noindent
where $\sigma_1(\omega)$ is the real part of the in-plane optical conductivity.  
In principle $W(\Omega,T)$ depends on both the cut-off  
$\Omega$ and the temperature $T$. When $\Omega \to \infty$, the f-sumrule for the optical conductivity is recovered. In this case $W \propto n$, where $n$ is the carrier density for a continuous system and therefore it is independent of $T$. 
In a tight-binding model $W(\Omega,T)$ is instead related to the carrier kinetic energy,\cite{mandague} and then retains a temperature dependence. 
In this framework, the behavior of $W(\Omega,T)$ provides crucial information on the energetic balance at $T_c$. Indeed, it has been reported that the onset of superconductivity both in underdoped and in optimally doped Bi$_2$Sr$_2$CaCu$_2$O$_{8+y}$ (BSCCO) is associated with a gain in kinetic energy,\cite{vdm,bont,hirsh} as opposed to the slight kinetic energy loss  in
conventional BCS superconductors.
Moreover, restricted sum rules, where in Eq. (\ref {weight}) $\Omega \leq \omega_p$, with $\omega_p$ the plasma edge, have been studied experimentally in 
BSCCO,\cite{vdm,bont} YBa$_2$Cu$_3$O$_{7-y}$ (YBCO),\cite{homes} and La$_{2-x}$Sr$_x$CuO$_4$ (LSCO),\cite{michele} both in the normal and in the superconducting phases. In the latter work, it has been shown that the quadratic $T$-dependence 

\begin{equation}
W(\Omega,T) \simeq W(\Omega,0) - B(\Omega)T^2,
\label{quadra}
\end{equation}

\noindent
holds for any infrared cut-off $\Omega$ between $\sim 0.1 eV$ 
and $\sim$ 1 eV, both in LSCO and in a conventional metal like gold. In both systems, when $\Omega$ increases, the $T$-independent term obviously increases, 
while the thermal response $B(\Omega)$ decreases. However, while in gold both $W(\Omega,0)$ and $B$ are controlled by the same energy scale (a hopping rate $t$), in LSCO the former scale is by one order of magnitude larger than the latter one.
Moreover, in LSCO a sizable temperature dependence of $W(\Omega,T)$
is found unexpectedly even for $\Omega \sim \omega_p$, while in gold 
$B(\Omega \simeq \omega_p)$ = 0.

Expression (\ref{quadra}) is easily recovered, at least for large cut-offs, 
if electron-electron interactions are neglected.  
If, for the sake of simplicity, one models the Cu-O plane 
with an effective single-band tight-binding model with nearest-neighbor 
hopping $t$ only, one has $W(\Omega \simeq \omega_p,T) = - (\pi e^2)/(2 
 d )\, E_{kin}$, where $e$ is the electric charge,  $d$ is the dimensionality of the
system and   $E_{kin}$ is the carrier kinetic energy. For a non-interacting system we can write $E_{kin}$ as an integral over the density of states (DOS) and  expand it {\it \`a la} Sommerfeld, obtaining Eq. (\ref{quadra}) with $W(\omega_p) \propto E_{kin}(T=0)$. Under these simplifying assumptions, 
$E_{kin}(T=0) \propto t$ and $B \propto t^{-1}$ (for a quasi-flat DOS, e.g., 
$B=\pi^2 e^2 /(24 \, t \, d)$), so that both quantities are controlled by the same energy scale $t$. As long as electronic correlations are neglected, this
result cannot be significantly changed by more realistic 
models. It has been recently shown that the addition of a next-nearest neighbor hopping $t'$ which controls the position of the Van Hove singularity cannot explain the observations in LSCO or BSCCO.\cite{lara}

The above standard behavior of non-interacting systems, in which the same
$t$ controls both $W(\Omega,0)$ and $B$, is manifestly at variance 
with experimental evidence in cuprates. 
In particular, in LSCO,\cite{michele} $E_{kin}(T=0)$ leads to
a $t_0$ of the order of several hundreds meV, 
consistent with independent estimates of the bandwidth based on photoemission 
spectra or band-structure calculations.\cite{pavarini}
On the other hand, the experimentally determined value of $B$ is unexpectedly
large and leads to $t_T \simeq$ 20 meV,\cite{michele,lara} smaller than $t_0$ by one order of magnitude. 
As it was proposed in Ref. \onlinecite{michele}, the existence of two distinct energy scales points towards the separation between low- and high-energy physics that is characteristic
of strongly correlated electronic systems, and suggests that the large difference
between $t_0$ and $t_T$ can be due to the proximity to a Mott insulator.

The present paper is aimed at putting this hint on firmer grounds by calculating
the optical conductivity of a strongly correlated single-band Hubbard model, by Dynamical Mean-Field Theory (DMFT).\cite{revdmft} We are aware that this approach will not provide a complete description of the optical properties of cuprates,
where spatial correlations and other effects definitely play a role.
However, we want to understand whether the Coulomb repulsion, even in its simplest
reliable description, provides the basic ingredient 
to explain the one-order-of-magnitude discrepancy between
the $T=0$ spectral weight and the thermal coefficient $B$. We will assume reasonable values for the parameters of the calculation, but in this context
it would be of little significance to finely tune them in order
to fit the experimental results. Those values, indeed, should 
be certainly modified as other effects beyond DMFT were included.

The Hamiltonian of the Hubbard model reads

\begin{eqnarray}
\label{hubbard}
{\cal H} &=& -t \sum_{<ij>\sigma} c_{i\sigma}^{\dagger} c_{j\sigma} 
+U\sum_{i}\left ( n_{i\uparrow}-{1\over 2}\right )
\left ( n_{i\downarrow}-{1\over 2}\right )+\nonumber\\
& & -\mu\sum_i (n_{i\uparrow}+n_{i\downarrow}),
\end{eqnarray} 
where $c_{i\sigma}(c^{\dagger}_{i\sigma})$ are annihilation(creation) operators
for fermions of spin $\sigma$ on site i, $n_{i\sigma} = c^{\dagger}_{i\sigma}
c_{i\sigma}$, $t$ is the hopping amplitude, $U$ is the local Hubbard repulsion,
 $\mu$ is the chemical potential and the sum in the first term is restricted to nearest-neighbors
only.

In order to solve  (\ref{hubbard}) we resort to the DMFT, presently the most reliable tool to treat the physics of doped Mott insulators. 
DMFT is based on the neglect of  spatial fluctuations,
but fully retains local quantum dynamics, and becomes exact in the
infinite coordination limit in which the self-energy
is local, i.e., momentum-independent.\cite{revdmft} 
Despite deliberately excluding phases with given spatial arrangement, DMFT 
allows for the simultaneous description of low- and high-energy features,
and has provided the first unified treatment of the correlation- and
density-driven Mott transitions. 
The DMFT maps the lattice model onto an impurity model subject to a 
self-consistency equation which contains the information about the original 
lattice. Here we consider, as it is often done, an infinite-coordination
Bethe lattice with a semi-circular DOS of half-bandwidth $D$. 
The impurity model has been solved by Exact 
Diagonalization at finite temperatures. The number of levels in the bath has 
been fixed to $N_b =4\div 5$, which, despite being small, gives a good description
of thermodynamic properties, at least at intermediate 
temperatures.\cite{revdmft} 
Following  Ref. \onlinecite{arun_qmc}, we assume  $D = 4t = 1.2 $eV,  and 
$U = 3D = 12 t$, so that  the antiferromagnetic superexchange 
$J = 4t^2/U \simeq 100$ meV is compatible with experimental estimates. 
As we discuss in the following, our choice is related to the message we aim
to bring with this paper. 

Before turning to the analysis of optical properties, 
we briefly remind how the strongly correlated phases arising by doping 
a Mott insulator are characterized
within DMFT. For large enough $U/D$, the half-filled system
is a Mott insulator, with a spectrum made by two broad incoherent 
features, the Hubbard bands, which are centered around $\pm U/2$
with a width of the order of the bare bandwidth.
As the system is doped, a quasiparticle (QP) peak appears around the Fermi level, whose weight is measured by the residue $Z = (1-\partial\Sigma(\omega)/\partial\omega\vert_{\omega =0})^{-1}$,
$\Sigma(\omega)$ being the (momentum-independent) self-energy.
$Z$, which tends to vanish when the Mott transition is approached, is
small in the strongly correlated regime, 
reflecting the loss of metallic behavior due to the Coulomb repulsion.
In the hole-doping case which we discuss here, the QP peak is located close
to the lower Hubbard band.

The knowledge of the single-particle spectral function can be directly 
used to compute the 
optical conductivity (and the associated spectral weight) because
vertex corrections vanish
in the infinite coordination limit in which DMFT becomes exact,\cite{revdmft} 
and $\sigma_1(\omega)$ (summed over all the spatial directions) 
is simply given by

\beqa
\sigma_1(\omega) & = & 
\frac{2 e^2}{\pi} \int \, d\e \, \int  \, d\nu \, \,
N(\e) V(\e)\mbox{Im}[ G(\e,\nu)] \nonumber \\
& \times  &  \mbox{Im}[G(\e,\nu+\omega)] 
\frac{f(\nu)-f(\nu+\omega)}{\omega}
\label{sig_DMFT}
\eeqa 
where $N(\e)$ is the DOS of the non-interacting system, 
and $V(\e)=(D^2-\e^2)/3$ is the square current
vertex of the infinite-coordination Bethe lattice we are 
considering,\cite{chatt} and $G(\e,\omega)=(\omega -\e +
\mu - \Sigma(\omega))^{-1}$ is the retarded
Green functions of the lattice system.

The behavior of $\sigma_1(\omega)$, already discussed in Refs. 
\onlinecite{jarrel,marcelo}, clearly mirrors the above described
spectral features of a doped Mott insulator, and presents three main
features, namely
(i) a Drude contribution at low frequencies (say up to $0.1 D$) arising from 
optical excitations  within the QP resonance, (ii)  a 
Mid Infrared (MIR) peak (at $\omega \sim 0.5 \div 1.0 D$) 
resulting from the transitions between  the lower Hubbard band and the unoccupied 
part of the quasi-particle peak and, finally, (iii) an
insulating-like contribution at higher frequencies (at $\omega \sim U$)
mainly due  to transitions between the Hubbard bands. 

Now we discuss how the above described behavior reflects in the temperature dependence of the 
optical spectral weight $W(\Omega,T)$. As it has been done for experimental data, 
we consider two values of the cut-off:  $\Omega = 0.1 D$, which is expected to include in the
integral only the contribution from the Drude peak, and $\Omega = 1.5 D$, which is in a sense 
the plasma edge of our model, i.e., contains both the Drude weight and the MIR bands, 
excluding from the integral the small but finite high-energy contributions.

\begin{figure}[t!]
\begin{center}
\includegraphics[width=46mm,height=45mm, angle=0]{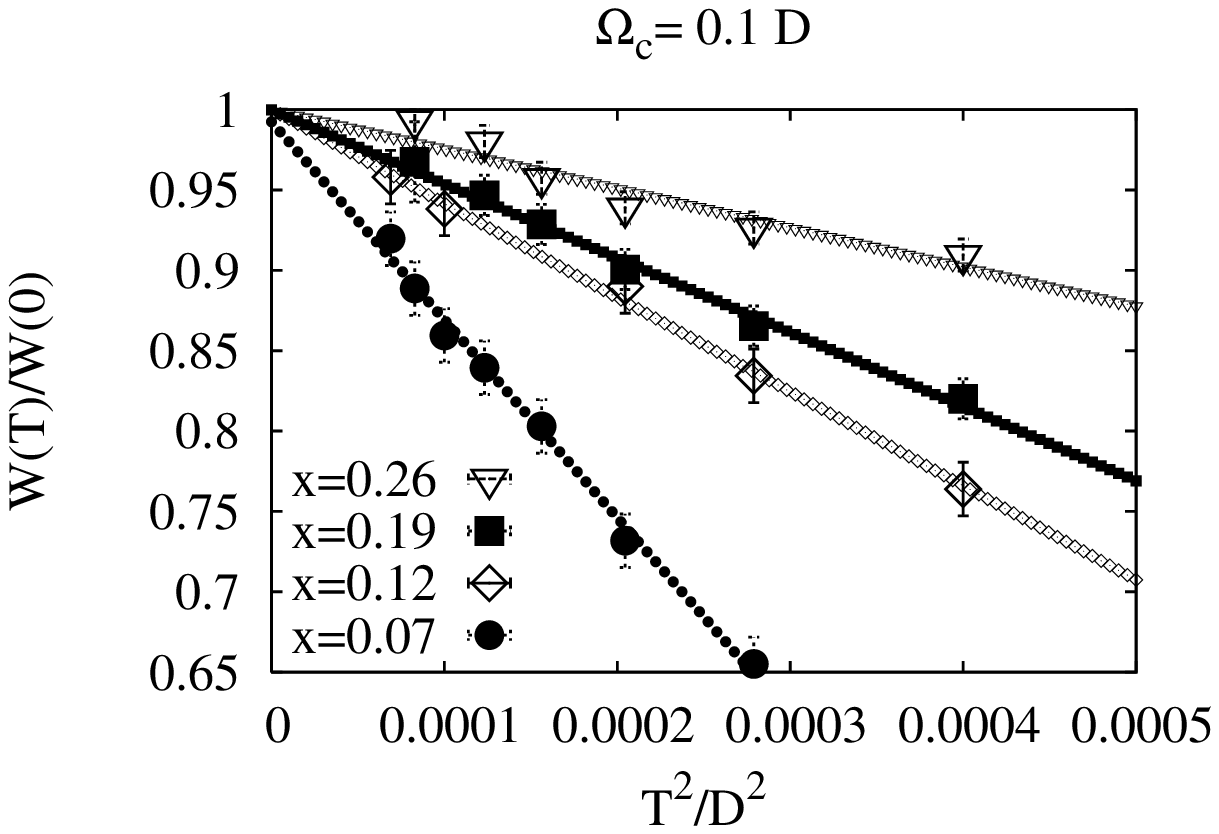}
\hspace{-7mm}
\includegraphics[width=45mm,height=45mm, angle=0]{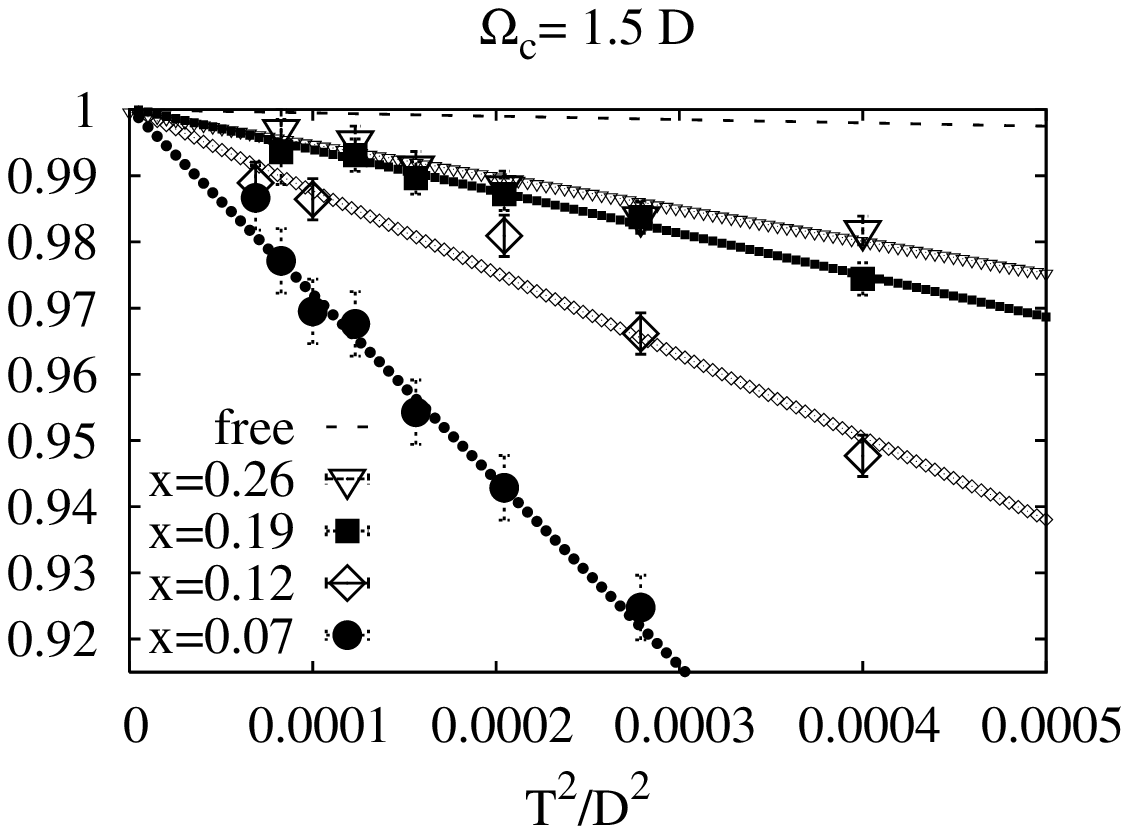}
\end{center}
\caption{\label{fig:slope_dmft}  $W(\Omega,T)$ as a function of $T^2$
for the Hubbard model, for $\Omega=0.1D$ (left panel)  and $\Omega=1.5D$ 
(right panel), normalized to its zero temperature extrapolation. Symbols are the results of the DMFT calculations, lines are best fits to them.}  
\end{figure}

In Fig. \ref{fig:slope_dmft} the ratio $W(T)/W(0)$ is plotted for both the above
$\Omega$ values and for different doping levels. One can see that our model calculation nicely follows the $T^2$ behavior in the range $0.008 D < T < 0.022 D$ 
(i.e., between 100 and 300K), as experimentally found in cuprates.\cite{vdm,bont,homes,michele}  The agreement with observations is not limited to the $T^2$ law, but extends to two important properties which are not accounted for by non-interacting calculations. Namely, we find that 
{\it (i)} the inclusion of strong correlations determines a substantial 
increase in the coefficient $B$ controlling the thermal excitations, and that
{\it (ii)} the  enhancement is  more pronounced for the lower value of $\Omega$.
The results for $B(\Omega = 0.1D, 1.5 D)$ and $W_0(\Omega) = W(\Omega,T=0)$ determined by fitting the numerical data with Eq. (\ref{quadra}) are reported in Table I. For the sake of comparison we also report the $T=0$ values of 
the quasiparticle weight $Z$ and for the full kinetic energy $E_{kin}$, i.e.,
$W(\Omega \to\infty, T=0)$.  

The large value of $B$ reproduced by the DMFT can be mainly ascribed
to the renormalized effective bandwidth $2D^* =2ZD$ determined by correlations.\cite{notamorethanb}
If however the same effective bandwidth controlled also the zero-temperature 
spectral weight in our calculation, 
a strong inconsistency between theory and experiment would appear for this last 
quantity, whose experimental value is 
compatible with photoemission estimates of the hopping parameter, in turn 
of the order of the bare hopping.

\begin{table}
\begin{center}
\begin{tabular}{||c|c|c|c|c|c|c||} \hline \hline
         &  $ B(0.1 D)$   & $B(1.5 D)$  & $W_0(0.1D)$  & $W_0(1.5D)$  & -$E_{kin}$    & $Z$   \\ \hline \hline 
x=0.07    &  $58\pm 5 $ & $27 \pm 3$ & $0.05 D$ &    $0.11 D$   &  $0.17 D$  & 0.12    \\ \hline
x=0.12    &   $48 \pm 5$  & $16 \pm 4$ & $0.08 D$ & $0.16 D$ & $0.21 D$  &   0.20    \\ \hline 
x=0.15    &   $43 \pm 5$  & $15 \pm 4$ & $0.09 D$ & $0.18 D$ & $0.22 D$  &   0.22    \\ \hline
x=0.19    &   $51 \pm 5 $  &  $13 \pm 5$ & $0.11 D$ & $0.20 D$  &  $0.24 D$ &  0.27   \\ \hline
x=0.26    & $33 \pm 7$ &  $ 11 \pm 3$  &  $0.14 D$ & $0.24 D$  & $0.27 D$  & 0.35  \\ \hline \hline
\end{tabular}
\caption{Results of the fits in Fig. 1 for selected doping values. First and second columns: values of the coefficient $B$ (in units of 
$1/D$) for low- and high-frequency cut-off, respectively (note that, for $U=0$, $B \simeq 2/D$ in the whole doping range here considered).  Third and fourth columns:  corresponding spectral weights.  Last two columns: kinetic energy and quasiparticle weight.}
\end{center}
\end{table}

\begin{figure}[t!]
\begin{center}
\includegraphics[width=80mm, angle=0]{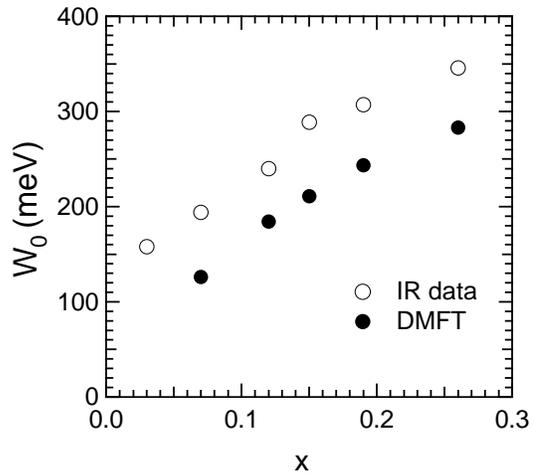}
\end{center}
\caption{\label{fig2}
Zero-temperature spectral weight for $\Omega = 1.5 D$ as a function of doping, for both the strongly correlated Hubbard model (dots) and for the observations in LSCO (circles). The latter data  are obtained by integrating the $\sigma_1(\omega)$ of LSCO in Ref. \onlinecite{lucarelli}.}
\end{figure}

The explicit calculation of $W_0(\Omega = 1.5 D)$
is instead in very good agreement with the experiments, as shown in Fig.\ \ref{fig2}. 
Therein the experimental points are obtained by integrating the $\sigma_1(\omega)$ of LSCO 
reported for different doping in Ref. \onlinecite{lucarelli}. Fig.\ \ref{fig2} shows that, even if the coherent hopping 
scale is significantly reduced by correlation, $W_0$ is much less affected and it is still large due to the 
contribution of the MIR band. Indeed it involves incoherent hopping processes 
which contribute to the average kinetic energy, even if they have no role in QP transport.

It is worth to notice that theoretical and experimental points clearly follow 
a similar behavior as a 
function of doping, with a decreasing weight when the Mott insulator is 
approached.

In this regard, we expect that the 
inclusion of spatial correlations beyond DMFT and of the antiferromagnetic
superexchange $J$ would result in a non vanishing renormalized
bandwidth, as predicted, e.g., by slave boson mean-field theories.\cite{marcogabi} 
At the same time, once the DMFT approximation is relaxed,
vertex corrections have to be considered in the optical conductivity.
The fact that in Fig. \ref{fig2} the calculated values are systematically lower
 than the experimental ones reflects our choice to fix $U/D$ without adjusting its value to data. 
Anyway, the ability of the present DMFT calculation to give the correct order of magnitude of $W_0$ 
suggest that the effect of the superexchange and
the inclusion of vertex correction may partially cancel each other, 
as it indeed happens within zero-temperature slave boson approaches.\cite{marcogabi}

\begin{figure}[t!]
\begin{center}
\includegraphics[width=80mm, angle=0]{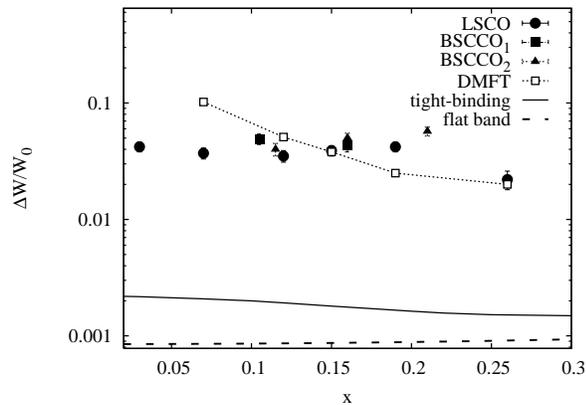}
\end{center}
\caption{\label{fig3}
The relative variation of the spectral weight between the lowest $T$ and 300 K, 
as a function of doping for various cuprates (full symbols) is compared with present DMFT calculations (squares) and
with the predictions of non-interacting models (tight binding model (boh line), flat band (boh-boh line)). The dashed line is a guide to the eye. The simple inclusion of correlation effects allows one to reproduce the observed absolute values with no need for fitting parameters. Data for LSCO are obtained from Refs. \onlinecite{michele,lucarelli}, for BSCCO from Ref. \onlinecite{bont} (BSCCO$_1$) and from Ref. \onlinecite{vdm} (BSSCO$_2$).} 
\end{figure}

In the final part of this paper we like to consider a quantity suitable to measure the variation with temperature of the optical spectral weight at large $\Omega$.
Denoting here as $W(T)$ the spectral weight $W(\Omega,T)$ where $\Omega = \omega_p$ for the experiments and $\Omega =1.5 D$ for the calculation, we plot in Fig.\ \ref{fig3} the relative variation of 
$W(T)$ between $T=0$ and $T$ = 300 K $\Delta W/W_0 = [W(300K) - W(0)]/W(0)$. In the same Figure, we report for comparison the prediction of a non interacting 
nearest neighbor model  (dotted line, see also Ref. \onlinecite{lara}). 
Fig.\ \ref{fig3} clearly shows that: i) such ratio has similar values in different families of cuprates, indicating 
that this quantity is  essentially material-independent; ii)
the thermal change of $W$ does not depend on details of the Fermi surface, which are instead dependent on both material and doping; iii) DMFT, with the introduction of strong correlation effects, reconciles the calculations with the experimental infrared behavior of the cuprates, which in no way can be reproduced by more conventional hopping models.

In conclusion, we have shown that the inclusion of strong correlation 
effects, already in the 
simplest theoretical scheme given by the Hubbard model, removes the 
inconsistency of non-interacting
models and naturally accounts for the bifurcation between a $T=0$ spectral 
weight which becomes rapidly comparable with the non-interacting value, 
and the coefficient of its  temperature dependence
 where a renormalized effective hopping plays the  major role.
This neither means that the Hubbard model in the DMFT
approximation can account for all properties of cuprates, nor that
our analysis has the ambition to be quantitative.
However we believe that a non-perturbative treatment of correlations,
such as the DMFT approach, is a basic building block for a theory 
of the optical response and of the transport properties of high-$T_c$ 
materials.
In our view, the next step would be the inclusion of spatial correlations
by means of non-local extensions of the DMFT such as Dynamical Cluster
Approximation (DCA)\cite{dca} or Cellular DMFT.\cite{cdmft} 
In the latter approach, the effective bandwidth is no
longer bound to vanish at the Mott transition, as it should also
depend on the antiferromagnetic exchange coupling $J$.
As mentioned above, we finally expect that the effect of the band broadening
induced by $J$ will be partially compensated by vertex corrections to the
optical conductivity.\cite{millis1,millis2}
These improvements would ultimately lead to a better understanding of the
role played by  charge and spin modulations in the optical response of cuprates. 

We acknowledge financial support from MIUR Cofin 2003.
We thank S. Biermann, A. Georges, M. Grilli, G. Sangiovanni and
 J.M. Tomczak for precious 
discussions, and L. Benfatto for constant support, useful suggestions and a 
careful reading of the manuscript.


\begin{thebibliography}{99}
\bibitem{vdm} H.J.A. Molegraaf {\it et al.},Science {\bf 295}, 2239 (2002).

\bibitem{bont} A.F. Santander-Syro {\it et al.}, Phys. Rev. Lett. 
{\bf 88}, 097005 (2002); A.F. Santander-Syro, R.P.S.M. Lobo and N. Bontemps, Europhys. Lett. {\bf 62}, 568 (2003); A. F. Santander-Syro, {\it et al.} Phys. Rev. B {\bf 70}, 134504 (2004).

\bibitem{homes} C.C. Homes {\it et al.} Phys. Rev. B {\bf 69}, 024514 (2004).

\bibitem{boris} A.V. Boris {\it et al.}, Science {\bf 304}, 708 (2004).

\bibitem{michele} M. Ortolani, P. Calvani, and S. Lupi,  Phys. Rev. Lett. {\bf 94}, 067002 (2005).

\bibitem{hirsh} J.E. Hirsch, Science, {\bf 295}, 2226 (2002).

\bibitem{norman}  M.R. Norman and  C. Pepin,  
Phys. Rev. B {\bf 66}, 100506(R) (2002).

\bibitem{millis1} A. J. Millis and H. D. Drew, Phys. Rev. B {\bf 67}, 214517 (2003). 

\bibitem{carbotte}  J.P. Carbotte and E. Schachinger, Phys. Rev. B {\bf 69}, 224501 (2004)

\bibitem{lara}  L. Benfatto, S. Sharapov, N. Andrenacci, and H. Beck, 
Phys. Rev. B in press (cond-mat/0407443).

\bibitem{millis2}  A. J. Millis, A. Zimmers, R. P. S. M. Lobo, and N. Bontemps
, cond-mat/0411172.  

\bibitem{mandague} P.F. Mandague, Phys. Rev. B {\bf 16}, 2437 (1977).

\bibitem{pavarini} E. Pavarini, I. Dasgupta, T. Saha-Dasgupta, O. Jepsen, 
and O.K. Andersen  Phys. Rev. Lett., {\bf 87}, 047003 (2001).

\bibitem{revdmft} A. Georges, G. Kotliar, M. J. Rozenberg, and W. Krauth, 
Rev. Mod. Phys. {\bf 68}, 13 (1996).

\bibitem{arun_qmc} A. Paramekanti, M. Randeria, and N. Trivedi,
Phys. Rev. Lett. {\bf 87}, 217002 (2001). 

\bibitem{chatt} A. Chattopadhyay, A.J. Millis and S. Sarma, Phys. Rev. B {\bf 61}, 10738 (2000).

\bibitem{jarrel} M. Jarrell, J.K. Freericks, and T. Pruschke, Phys. Rev. B
 {\bf 51}, 11704 (1995).

\bibitem{marcelo} M. J. Rozenberg, {\it et al.}, 
Phys. Rev. Lett. {\bf 75}, 105 (1995)


\bibitem{lucarelli} A. Lucarelli {\it et al.}, Phys. Rev. Lett. {\bf 90}, 037002 (2003).


\bibitem{marcogabi} M. Grilli and G. Kotliar, Phys. Rev. Lett. {\bf 64}, 1170 (1990).

\bibitem{dca} M. H. Hettler {\it et al.}, Phys. Rev. B 58, R7475 (1998)

\bibitem{cdmft} G. Kotliar, S. Savrasov, G. Palsson, and G. Biroli, 
Phys. Rev. Lett. {\bf 87}, 186401 (2001)

\bibitem{notamorethanb} We notice that the enhnacement of $B$ is larger than
$1/Z$, suggesting a possible role for finite lifetime effects.
\end{thebibliography}
\end{document}